# Brownian Motion and Quantum Mechanics

Roumen Tsekov

Department of Physical Chemistry, University of Sofia, 1164 Sofia, Bulgaria

A theoretical parallel between the classical Brownian motion and quantum mechanics is explored. It is shown that, in contrast to the classical Langevin force, quantum mechanics is driven by turbulent velocity fluctuations with diffusive behavior. In the case of simultaneous action of the two stochastic sources, the quantum Brownian motion takes place, which is theoretically described as well.

Because the Schrödinger equation is a parabolic partial differential one, its similarity to diffusion puzzles scientists from the inception of quantum mechanics. In a seminal paper, Nelson[1] succeeded to derive the Schrödinger equation from fundamental diffusion. His derivation suffers, however, a major shortcoming - the drift term in the key stochastic equation depends on the probability density, which indicates a mean field approach. In the present study, we are trying to eliminate this defect by deriving the Schrödinger equation from first principles for an arbitrary system of quantum particles. To keep the presentation transparent, similar to Nelson we do a parallel to the classical Brownian motion at every step.

Introducing the $3N$-dimensional vectors of the positions $X(t)$ and momenta $P(t)$ of $N$ particles, their dynamics obeys the following stochastic Hamilton equations in the frameworks of the classical Brownian motion theory[2]

$$\dot{X} = M^{-1} \cdot P \qquad \dot{P} = F - \nabla U \qquad (1)$$

Here $M$ is a $3N \times 3N$ diagonal mass matrix, $U(X)$ is a general potential, $\nabla$ is a $3N$-dimensional positions nabla operator and $F$ is a $3N$-dimensional stochastic force. The latter originates from the surroundings and contains the friction force as well. It is important to note, that the classical Brownian motion is force-driven, i.e. the noise is active in the momentum subspace. From quantum mechanics, it is well known that a free quantum particle possesses constant momentum.

Therefore, the force $F$ is not present by definition in vacuum, which is a continuous phase without discrete structure. However, since the quantum particles are still randomly distributed in the position subspace, it follows that the quantum stochastic equations should read

$$\dot{X} = C + M^{-1} \cdot P \qquad \dot{P} = -\nabla U \qquad (2)$$

Here $C$ is a stochastic velocity, which is the source of randomness in quantum mechanics. One can imagine it as a stochastic wind in vacuum. Nonetheless, in both cases the Newtonian dynamics is given by the stochastic Langevin equation[3]

$$\ddot{X} = A - M^{-1} \cdot \nabla U \qquad (3)$$

where the statistical properties of the $3N$-dimensional stochastic acceleration $A = M^{-1} \cdot F$ or $A = \dot{C}$ solely discriminates between the Brownian and quantum motions, respectively.

Let us introduce the $3N$-dimensional local density $\rho \equiv \delta^{3N}(x - X)$ of the particles in the configurational space. Differentiating $\rho$ on time yields the continuity equation

$$\partial_t \rho + \nabla \cdot (\rho \dot{X}) = 0 \qquad (4)$$

Another differentiation on time of the particles flow $\rho \dot{X}$ yields in combination with Eq. (3) the momentum balance in stochastic hydrodynamic-like form

$$\partial_t (\rho \dot{X}) + \nabla \cdot (\rho \dot{X} \dot{X}) = \rho (A - M^{-1} \cdot \nabla U) \qquad (5)$$

The random fluctuations of $A$ are supposed to be very fast and irregular. For this reason, one can average the equations above either on time or on an ensemble of realizations to obtain

$$\partial_t \overline{\rho} + \nabla \cdot (\overline{\rho} V) = 0 \qquad \partial_t (\overline{\rho} V) + \nabla \cdot (\overline{\rho} V V + \overline{\rho} R) = \overline{\rho A} - \overline{\rho} M^{-1} \cdot \nabla U \qquad (6)$$

Here $V \equiv \overline{\rho \dot{X}} / \overline{\rho}$ is the local-mean velocity, while $R \equiv \overline{\rho (\dot{X} - V)(\dot{X} - V)} / \overline{\rho}$ is the $3N \times 3N$-dimensional Reynolds stress, accounting for the stochastic energy in general. The further use of Eq. (6) requires modeling of the unknown Favre-averaged statistical moments $R$ and $\overline{\rho A}$.

Let us consider first the classical Brownian motion, where the picture is well known. The Reynolds stress $R = k_B T M^{-1}$ is constant proportional to temperature. The local-mean force from the surroundings is the friction force $\overline{\rho F} / \overline{\rho} = -B \cdot V$, where $B$ is the $3N \times 3N$ friction coefficients matrix. Introducing these quantities in Eq. (6), the latter changes after some rearrangements to

$$M \cdot (\partial_t V + V \cdot \nabla V) + B \cdot V = -\nabla(U + k_B T \ln \overline{\rho}) \qquad (7)$$

At strong friction one can neglect the first inertial term as compared to the second one and derive the local-mean velocity $V = -B^{-1} \cdot \nabla(U + k_B T \ln \overline{\rho})$. Substituting the latter into the continuity equation (6) yields the well-known Smoluchowski equation[2]

$$\partial_t \overline{\rho} = \nabla \cdot B^{-1} \cdot (\overline{\rho} \nabla U + k_B T \nabla \overline{\rho}) \qquad (8)$$

It is easy to recognize here the Einstein diffusion matrix $k_B T B^{-1}$. Due to the dependence of the friction force on the velocity $V$, the classical diffusion is an irreversible process.

The situation in quantum mechanics is different. The Schrödinger equation is reversible due to the lack of friction. Since according to Eq. (2) there is no random mixing in the momentum space, the Reynolds stress reduces to a dyadic form[4] $R = WW$, where $W$ is the local-mean-square image of the stochastic velocity $C$. Assuming diffusive origin of the latter one can write

$$\overline{\rho} W = -D \cdot \nabla \overline{\rho} \qquad (9)$$

Hence, the turbulent flow is represented by a Fick flux with a $3N \times 3N$ diffusion matrix $D$, which is not identified yet. The other unknown term in Eq. (6) can be specified consecutively to

$$\overline{\rho A} = \overline{\rho \dot{C}} = -\overline{\dot{\rho} C} = \nabla \cdot \overline{\rho \dot{X} C} = \nabla \cdot [-D \cdot \nabla(\overline{\rho} W)] = \nabla \cdot [D \cdot \nabla(D \cdot \nabla \overline{\rho})] \qquad (10)$$

As expected, it is a linear functional of $\overline{\rho}$.[5] In general, the independence of all terms above from the velocity $V$ ensures the time-reversibility of quantum mechanics. Introducing Eqs. (9) and (10) in Eq. (6) yields after rearrangements the form

$$\partial_t \overline{\rho} + \nabla \cdot (\overline{\rho} V) = 0 \qquad \partial_t V + V \cdot \nabla V = -M^{-1} \cdot \nabla U - \nabla \cdot (\overline{\rho} D \cdot \nabla W)/\overline{\rho} \qquad (11)$$

One can recognize in the last term in the brackets the turbulent viscous stress with the kinematic turbulent viscosity matrix $D$, which describes naturally the turbulent diffusion of the turbulent velocity as well. Substituting the turbulent velocity $W = -D \cdot \nabla \ln \overline{\rho}$ in Eq. (11) yields after serious rearrangements the closed form

$$\partial_t V + V \cdot \nabla V = -M^{-1} \cdot \nabla U + 2D \cdot \nabla(\nabla \cdot D \cdot \nabla \overline{\rho}^{1/2}/\overline{\rho}^{1/2}) \qquad (12)$$

After Nelson[1] the universal diffusion matrix reads $D \equiv \hbar M^{-1}/2$. Introducing it in Eq. (12) the latter changes to

$$\partial_t \overline{\rho} + \nabla \cdot (\overline{\rho} V) = 0 \qquad M \cdot (\partial_t V + V \cdot \nabla V) = -\nabla(U + Q) \qquad (13)$$

where $Q \equiv -\hbar^2 \nabla \cdot M^{-1} \cdot \nabla \overline{\rho}^{1/2}/2\overline{\rho}^{1/2}$ is the Bohm quantum potential.[6] Introducing the local-mean velocity potential via $\nabla S \equiv M \cdot V$ and the wave function $\psi \equiv \overline{\rho}^{1/2} \exp(iS/\hbar)$, Eq. (13) reduces straightforward to the Schrödinger equation for arbitrary $N$ particles

$$i\hbar\partial_t\psi = -\hbar^2\nabla\cdot M^{-1}\cdot\nabla\psi/2 + U\psi = \hat{H}\psi \qquad (14)$$

where $\hat{H}$ is the standard Hamiltonian operator.

Finally, one can analyze the quantum Brownian motion, where both stochastic sources are simultaneously acting

$$\dot{X} = C + M^{-1}\cdot P \qquad\qquad \dot{P} = F - \nabla U \qquad (15)$$

If the two noises do not interfere, one can write straightforward a joint version of Eq. (6). However, since the quantum potential becomes temperature-dependent via the probability density, it shows that the thermal fluctuations are accounted twice.[7] To subtract the entropic effect from the quantum potential, it should be replaced by the corresponding free energy $k_B T\int Q d\beta$, calculated via the Gibbs-Helmholtz relation, where $\beta\equiv 1/k_B T$ is the reciprocal temperature. It coincides naturally with $Q$ at zero temperature. Thus, the correct joint form of Eq. (6) reads

$$\partial_t\bar{\rho} + \nabla\cdot(\bar{\rho}V) = 0 \qquad\qquad M\cdot(\partial_t V + V\cdot\nabla V) + B\cdot V = -\nabla(U + k_B T\int Q d\beta + k_B T\ln\bar{\rho}) \qquad (16)$$

At strong friction, the local-mean velocity $V = -k_B T B^{-1}\cdot\nabla[\int(U+Q)d\beta + \ln\bar{\rho}]$ can be directly expressed from the second one and introducing it into the first continuity equation yields the quantum Smoluchowski equation[8]

$$\partial_t\bar{\rho} = \nabla\cdot k_B T B^{-1}\cdot[\bar{\rho}\nabla\int\bar{\rho}^{-1/2}(\hat{H} + 2\partial_\beta)\bar{\rho}^{1/2}d\beta] \qquad (17)$$

In this equation, the detailed form of the quantum potential is applied. The integral above represents the full free energy functional, while the integrated quantity is the energy of the whole system. The equilibrium solution of Eq. (17) does not depend on time and satisfies the following temperature-dependent Schrödinger equation

$$-2\partial_\beta (Z\bar{\rho}_{eq})^{1/2} = \hat{H}(Z\bar{\rho}_{eq})^{1/2} \tag{18}$$

where $Z$ originates from the mean energy $E \equiv -\partial_\beta \ln Z$. This equation reduces to the ordinary Schrödinger equation after the half Wick rotation $\beta/2 \to it/\hbar$. Its solution of is the well-known quantum canonical Gibbs distribution[8]

$$Z\bar{\rho}_n = \exp(-\beta E_n)\psi_n^2 \tag{19}$$

where $\{E_n, \psi_n\}$ are the temperature independent eigenvalues and normalized eigenfunctions of the system Hamiltonian, $\hat{H}\psi_n = E_n\psi_n$. From the normalization of Eq. (19) it follows the well-known from quantum statistical thermodynamics expression $Z = \sum \exp(-\beta E_n)$ for the canonical partition function.

Although Eq. (17) is a complicate nonlinear differential equation, its solution of for free particles ($U \equiv 0$) is simply a zero-centered multivariable normal distribution density with the co-variance matrix $\Sigma$, which satisfies the following differential equation

$$2\beta B \cdot \Sigma \cdot \partial_t \Sigma^{-1} + \hbar^2 \int \Sigma^{-1} \cdot M^{-1} \cdot \Sigma^{-1} d\beta + 4\Sigma^{-1} = 0 \tag{20}$$

If one considers a diagonal friction matrix $B = \gamma M$, the solution $\Sigma = \eta M^{-1}$ of this equation is also diagonal, where the dispersion $\eta$ obeys the dynamic equation

$$2\beta\gamma\partial_t \eta = 4 + \eta \int (\hbar/\eta)^2 d\beta \tag{21}$$

The integral on the reciprocal temperature $\beta$ complicates the mathematical problem. However, either at low or at high temperature one can use the l'Hopital rule to approximate Eq. (21) to

$$2\gamma\partial_t \eta = 4/\beta + \hbar^2/\eta \tag{22}$$

It is now possible to integrate this equation and the result reads

$$\eta - (\hbar^2\beta/4)\ln(1 + 4\eta/\hbar^2\beta) = 2t/\gamma\beta \qquad (23)$$

The Einstein law $\Sigma = 2k_B T B^{-1} t$ follows from Eq. (23) in the classical limit, while at zero temperature the quantum sub-diffusive expression $\Sigma \cdot \Sigma = \hbar^2 M^{-1} \cdot B^{-1} t$ holds.